\input{aipcheck}

\documentclass[
    ,final            
  ]
  {aipproc}

\layoutstyle{6x9}

\begin{document}

\title{Modeling global event properties using hydrodynamics from RHIC to LHC}


\author{Piotr Bo\.zek}{
  address={The H. Niewodnicza\'nski Institute of Nuclear Physics,
PL-31342 Krak\'ow, Poland}
  ,altaddress={Institute of Physics, Rzesz\'ow University, 
PL-35959 Rzesz\'ow, Poland}}

\begin{abstract}
The relativistic hydrodynamic model is applied to describe the 
expansion of the dense matter formed in relativistic heavy-ion collisions.
The hydrodynamic expansion of the fluid, supplemented with the 
statistical emission 
of hadrons at freeze-out  gives a satisfactory description of the observables 
for particles emitted with soft momenta. Experimental data for the
transverse momentum spectra, elliptic flow and interferometry radii give
 constraints on the properties of the fluid, its equation of state and 
viscosity coefficients. The role of the fluctuations of the 
initial profile of the energy 
density is discussed.
\end{abstract}

\maketitle


\section{Introduction}

Experimental studies of heavy-ion collisions at ultrarelativistic energies 
have the aim at  creating a drop of very dense and hot matter  and  measuring 
its properties. Measurements at the BNL Relativistic Heavy Ion Collider (RHIC)
cover a range of energies up to $\sqrt{s_{NN}}=200$GeV for Cu-Cu and Au-Au
 systems. Recent results from the CERN Large Hadron Collider (LHC) extend 
the available energies to $\sqrt{s_{NN}}=2.76$TeV for Pb-Pb interactions.

The results accumulated in the last decade indicate that a fireball 
of strongly interacting, expanding fluid is formed 
\cite{Adcox:2004mh,Arsene:2004fa,Adams:2005dq,Back:2004je,Aamodt:2010pa}
in the collision. Observables related to the emission of particles with 
soft momenta demonstrate the onset of a strong collective flow of  matter.
Particle spectra are written using Fourier coefficients of the directed 
$v_1$, elliptic $v_2$, triangular  $v_3$, or higher harmonic flow
\begin{eqnarray}
\frac{dN}{d^2p_\perp dy}&=&\frac{dN}
{2\pi p_\perp dp_\perp dy}\left(1+v_1 \cos(\phi-\Psi_1)
+v_2 \cos\left(2(\phi-\Psi_2)\right) \right. \nonumber \\
& & \left. +v_3 \cos\left(3(\phi-\Psi_3)\right)+\dots \right) \ .
\end{eqnarray}
The flow coefficients can be extracted using the orientation angles $\Psi$
or from many-particle cummulants.
Transverse momentum spectra of identified particles can be quantitatively 
understood as coming from the convolution of the collective flow of the 
fluid elements and of the thermal emission 
\cite{Schnedermann:1993ws,Retiere:2003kf}. This observation
is
even more striking for the azimuthal asymmetry of momentum distributions, 
the elliptic flow. The observation of the elliptic flow has a very 
convincing interpretation in terms of the collective expansion of a 
source with an azimuthally asymmetric initial geometry \cite{Ollitrault:1992}.
Two-particle correlations of same particles allow to measure the 
interferometry radii of the emission region \cite{Wiedemann:1999qn}.
The values of the radii and their dependence on the momentum of 
the pion pair show the presence of a strong transverse flow. The extracted 
short emission time points towards a rapid expansion of the fireball.

We discuss the interpretation of the expansion of the fireball in terms of the 
perfect fluid 
\cite{Teaney:2000cw,Kolb:2003dz,Hirano:2002ds,Hama:2005dz,Broniowski:2008vp,Huovinen:2006jp,Bozek:2009ty}  and viscous hydrodynamics 
\cite{Teaney:2009qa,Romatschke:2009im,Song:2008hj,Bozek:2009dw,Ollitrault:2010tn,Schenke:2010rr}. Many efforts have been devoted to the construction of a model giving 
a realistic description  of the data. These studies yield a dynamical picture
of the 
 space-time evolution of the bulk of the matter created in the collision.
At the same time they give access to some physical properties of the system, the size and life-time of the fireball, the equation of state, the shear 
viscosity coefficient. Many unknowns that come into play in the construction
 of the model, require the use of 
 as many experimental constraints as possible to 
limit the possible systematic errors. In that way, a satisfactory description
 of the soft observables limits the effect of the model uncertainties or even 
helps to get some additional information, e.g on the role of the
 fluctuations in the initial state.

\section{Relativistic Hydrodynamics}

The expansion of the fireball is described using relativistic hydrodynamics.
The basic object in the formalism of second-order viscous hydrodynamics is
the energy-momentum tensor \cite{IS}
\begin{equation}
T^{\mu\nu}=(\epsilon+p)u^\mu u^\nu - p g^{\mu\nu} +\pi^{\mu\nu} +\Pi \Delta^{\mu\nu},
\end{equation}
the energy-momentum tensor of the perfect fluid gets a correction in the 
form of the stress tensor $\pi^{\mu\nu}$ from shear viscosity and
from bulk viscosity $\Pi$,
$u^\mu$ is the fluid velocity,  $\Delta^{\mu\nu}=g^{\mu\nu}-u^\mu u^\nu$.
We note, that  additional transport coefficients are possible in general
\cite{IS,Baier:2007ix}. The hydrodynamic equations
\begin{equation}
\partial_\mu T^{\mu\nu}=0
\end{equation}
and the equation of state connecting the pressure $p$ 
and the energy density $\epsilon$ are supplemented 
with dynamical equations for the  stress corrections
\begin{equation}
\Delta^{\mu \alpha} \Delta^{\nu \beta} u^\gamma \partial_\gamma \pi_{\alpha\beta}=\frac{2\eta \sigma^{\mu\nu}-\pi^{\mu\nu}}{\tau_{\pi}}-\frac{1}{2}\pi^{\mu\nu}\frac{\eta T}{\tau_\pi}\partial_\alpha\left(\frac{\tau_\pi u^\alpha}{\eta T}\right) 
\end{equation}
and
\begin{equation}
 u^\gamma \partial_\gamma \Pi=\frac{-\zeta \partial_\gamma u^\gamma-\Pi}{\tau_{\Pi}}-\frac{1}{2}\Pi\frac{\zeta T}{\tau_\Pi}\partial_\alpha\left(\frac{\tau_\Pi u^\alpha}{\zeta T}\right)  \ . \end{equation}
Two viscosity coefficients appear in the equations, 
the shear viscosity $\eta$ and the bulk viscosity $\zeta$ and two
  relaxation times $\tau_\pi$ and $\tau_\Pi$. The value of the viscosity coefficients
 can be estimated in kinetic theory if the cross sections are known.
However, for very strongly interacting systems kinetic theory breaks down
\cite{Danielewicz:1984ww}.
In  AdS/CFT calculations the ratio of the shear viscosity 
to entropy reaches a very small value $\eta/s=1/4\pi$ \cite{Kovtun:2004de}.
The small value of the shear viscosity, as inferred from the experimental data 
\cite{Adcox:2004mh,Arsene:2004fa,Adams:2005dq,Back:2004je,Aamodt:2010pa}, 
justifies the use of the perfect fluid hydrodynamics 
as a first approximation. 

Perfect fluid calculations can explain a large number of observations \cite{Teaney:2000cw,Kolb:2003dz,Hirano:2002ds,Hama:2005dz,Broniowski:2008vp,Huovinen:2006jp,Bozek:2009ty,Hirano:2005xf}.
Calculations have been performed both in simplified, boost-invariant $2+1$-D geometry 
as well as in  $3+1$-D.   The generation of the transverse flow is correctly described
as
the acceleration of the collective flow from pressure gradients in the source. 
The predicted elliptic flow overshoots  the experimental points, the calculation can be made closer 
to the data if a hadronic cascade stage is introduced after 
the hydrodynamic expansion \cite{Hirano:2005xf}.
The Hanbury Brown-Twiss  (HBT) correlations could not 
be described in early calculations using an equation of state with a first order phase transition 
\cite{Hirano:2002ds,Aguiar:2001ac,Morita:2003mj,Morita:2006zz}.
The agreement is greatly improved when using lattice QCD inspired parameterization of the equation of state with a cross over 
\cite{Broniowski:2008vp,Pratt:2008qv}.

To illustrate the quality of the predictions of perfect fluid hydrodynamics ,
we show an example of a $3+1$-D hydrodynamic simulation \cite{Bozek:2009ty}
 with a 
cross-over equation of state \cite{Chojnacki:2007jc}.
\begin{figure}
  \includegraphics[height=.45\textheight]{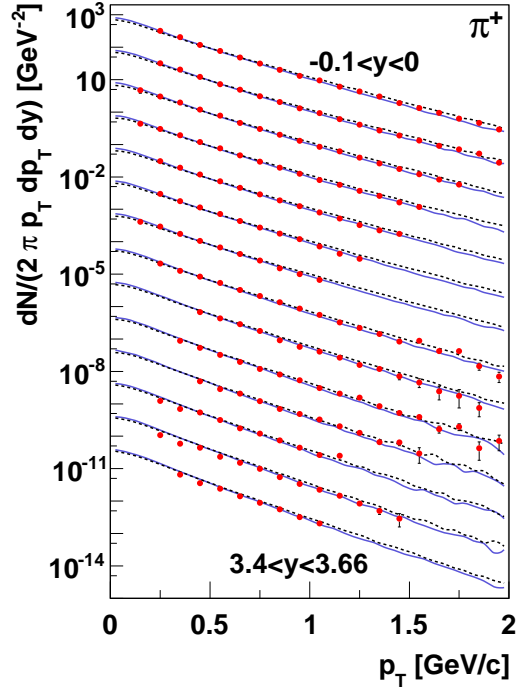}
  \caption{Transverse momentum spectra of pions produced in Au-Au collisions 
at $\sqrt{s}=200$GeV at different rapidities, compared to 
BRAHMS Collaboration Data
\cite{Arsene:2003yk} (from \cite{Bozek:2009ty})}
\label{fig:pions}
\end{figure}
In Fig. \ref{fig:pions} are shown results for the pion spectra at different 
rapidities. Spectra of pions and kaons can be well described 
for soft momenta $p_\perp<1.5$GeV in a perfect fluid $3+1$-D model. 
Up to now, there exist only one calculation of viscous hydrodynamics in $3+1$-D
\cite{Schenke:2010rr}, other viscous codes use the  boost-invariant geometry.

\begin{figure}
  \includegraphics[height=.5\textheight]{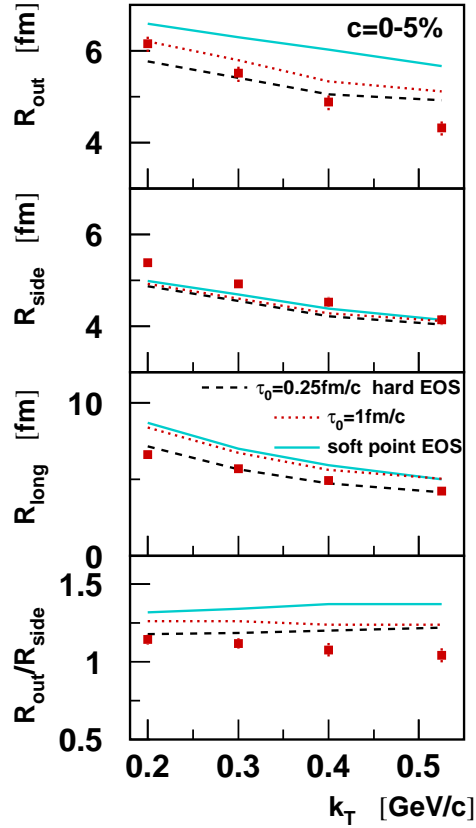}
  \caption{HBT radii for hydrodynamic calculation with a cross-over
transition (dashed lines) and for an equation of state with some softening 
around $T_c$ (solid lines)
 (from \cite{Bozek:2009ty}).}
\label{fig:hbtcompar}
\end{figure}

We illustrate the sensitivity of the HBT radii, and especially of the 
ratio $R_{out}/R_{side}$ to the softening of the equation of state. 
The increase of this ratio has been proposed as a measure of the 
strength of the first order transition \cite{Rischke:1996em}.
Experiments at RHIC and the LHC exclude such a behavior.
\begin{figure}
  \includegraphics[height=.22\textheight]{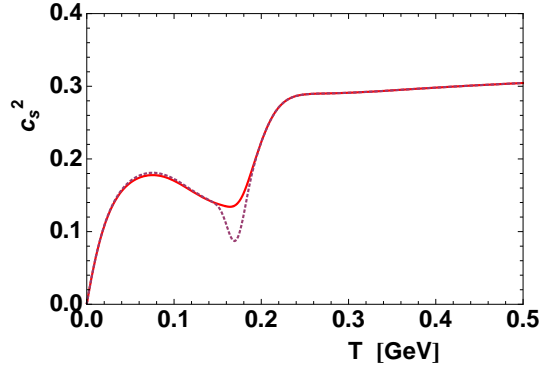}
  \caption{Temperature dependence of the square of the velocity of sound, for
a  cross-over equation of state (solid line)  \cite{Chojnacki:2007jc} 
and for an equation of state with  a softening around $T_c$ (dashed line)
 (from \cite{Bozek:2009ty})}
\label{fig:cs}
\end{figure}
When using the equation of state with 
minimal softening around $T_c$ (Fig. \ref{fig:cs}), one obtains the best agreement with the measured interferometry radii (Fig. \ref{fig:hbtcompar}).
The success of the simulations using a cross-over equation of state 
\cite{Broniowski:2008vp,Pratt:2008qv} is a phenomenological confirmation of 
recent lattice QCD calculations of the equation of state \cite{Aoki:2006we}.

\begin{figure}
  \includegraphics[height=.53\textheight]{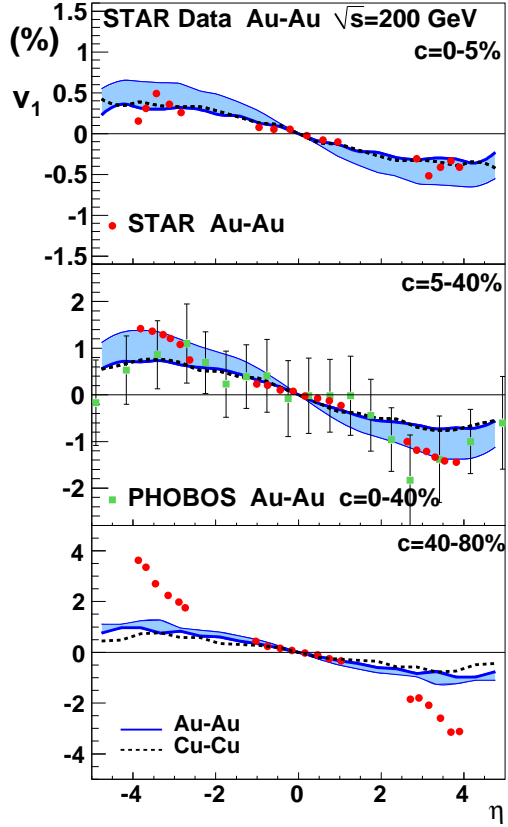}
  \caption{Directed flow of charged particles from $3+1$-D 
hydrodynamic calculation, compared to the 
data of STAR and PHOBOS Collaborations \cite{Back:2005pc,Abelev:2008jga}.
 (from \cite{Bozek:2010bi}). The color band indicate the systematic error form 
the range of the uncertainty in the initial tilt of the source.}
\label{fig:v1}
\end{figure}
The directed flow has been measured at ultrarelativistic energies. The 
formation of the flow in the hydrodynamic expansion implies a non-boost invariant 
geometry and requires a mechanism breaking the symmetry around the beam axis.
Asymmetric emission from forward and backward going participants 
\cite{Bialas:2004su} leads to a tilt of the source. The
 model  reproduces the observed value of the 
directed flow, in Au-Au and Cu-Cu collisions (Fig. \ref{fig:v1}).
The formation of the directed flow can serve as probe of the early 
non-equilibrium imbalance between the longitudinal and transverse pressures
\cite{Bozek:2010aj}. 

\begin{figure}
  \includegraphics[height=.23\textheight]{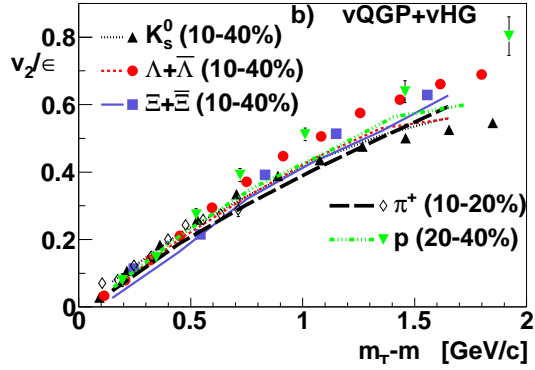}
  \caption{Elliptic flow of identified particles scaled by the initial eccentricity 
in hydrodynamic calculations compared to the data of PHENIX and STAR Collaborations \cite{Adams:2005zg},  \cite{Adler:2003kt},  \cite{Adams:2004bi}
 (from \cite{Bozek:2010er}).}
\label{fig:v2sc}
\end{figure}

The elliptic flow is generated in the expansion of an initial 
azimuthally deformed source. The final elliptic flow depends
 on the time of the evolution, on the viscosity coefficient
 \cite{Teaney:2003kp} and on the 
initial eccentricity. The measured elliptic flow can be used to extract the value of the shear viscosity coefficient
\cite{Luzum:2008cw,Song:2010mg}. The main uncertainty is 
related to the value of the initial eccentricity of the fireball. 
Depending on the assumed model of the initial density profile 
(Glauber model or KLN model) the extracted value of the shear viscosity
lies between $\eta/s=0.08$ and $\eta/s=0.2$. The calculation of the
 elliptic flow for identified particles requires a realistic description of the 
hadronic stage of the collision
\cite{Bozek:2010er,Song:2011qa}. If this stage is modelled by hydrodynamics, 
 bulk viscosity must be introduced. In Fig. \ref{fig:v2sc} 
the results for $v_2$ of different particles are shown using $\eta/s=0.1$ 
and $\zeta/s=0.04$  in the hadronic phase.
In summary, hydrodynamic studies of RHIC data have shown that a strongly
 interacting matter is formed in the collision, with a cross-over  equation
 of state and a small viscosity. The expansion of the the fireball is very 
rapid.

\section{From RHIC to LHC}

\begin{figure}
  \includegraphics[height=.27\textheight]{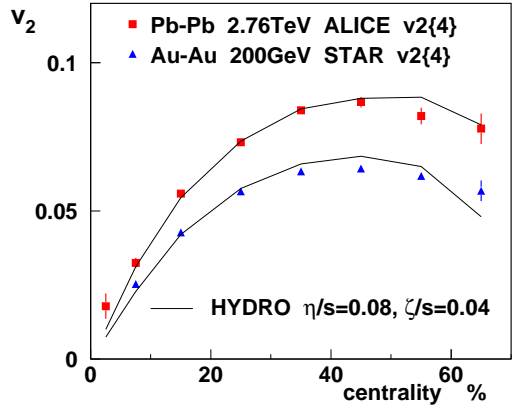}
  \caption{ Elliptic flow coefficient in Au-Au collisions at $\sqrt{s}=200$GeV
(STAR data \cite{Adams:2004bi}),
 and for Pb-Pb collisions at $2.76$TeV (ALICE data \cite{Aamodt:2010pa}),
 compared to viscous hydrodynamics results \cite{Bozek:2011wa}.}
\label{fig:alicev2}
\end{figure}
\begin{figure}
  \includegraphics[height=.3\textheight]{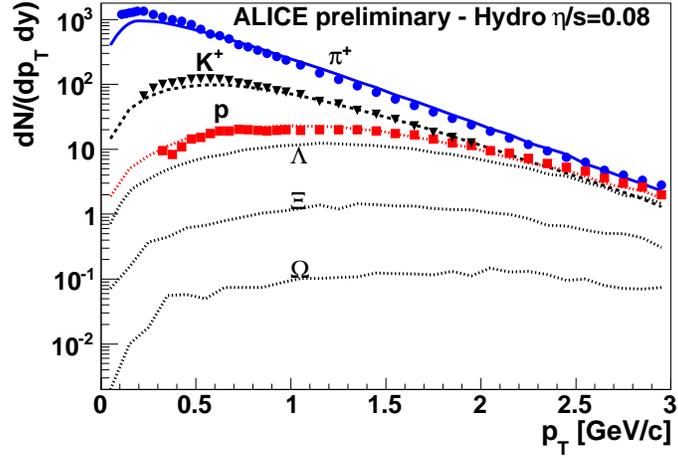}
  \caption{Transverse momentum spectra of identified particles from
 viscous hydrodynamics, compared to preliminary ALICE Collaboration 
data \cite{Floris:2011ru}.}
\label{fig:alicept}
\end{figure}
\begin{figure}
  \includegraphics[height=.5\textheight]{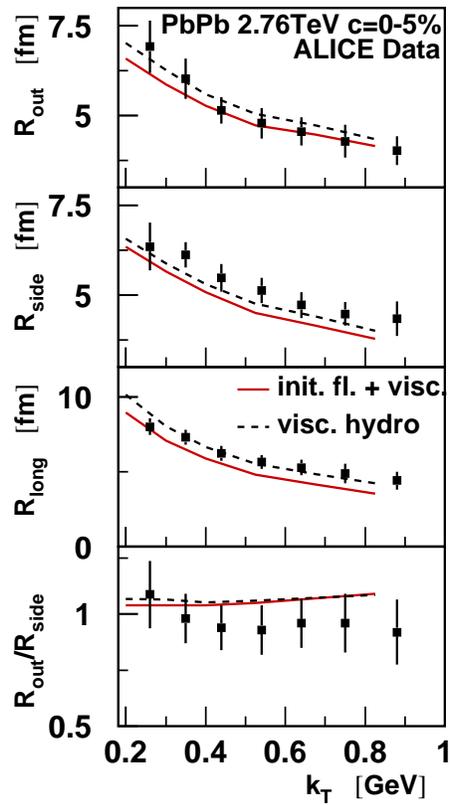}
  \caption{HBT  radii for pions emitted in Pb-Pb Collisions at
 $\sqrt{s}=2.76$GeV. The hydrodynamic model results are denoted  by the
 dashed lines for zero initial flow
 and by the solid lines when the initial pre-equilibrium flow is included 
\cite{Bozek:2010er},  ALICE Collaboration data
 \cite{Aamodt:2011mr} (from \cite{Bozek:2010er})}
\label{fig:hbtlhc}
\end{figure}
From the recent data on
 Pb-Pb collisions at the LHC \cite{Aamodt:2011mr,Aamodt:2010pa}
 a similar picture of the dynamics emerges as at RHIC energies.
The multiplicity and the life-time of the system increases, but the 
global properties are very similar.
In a first approach, the same parameters as 
used at $\sqrt{s}=200$GeV may be used ($\eta/s=0.08$ for Glauber initial conditions and $\zeta/s=0.04$ below $T_c$) 
increasing only the initial energy density.
Several calculation exist for LHC energies 
\cite{Luzum:2010ag,Schenke:2011tv,Shen:2011eg,Song:2011qa,Bozek:2011wa,Bozek:2010er,Karpenko:2011qn,Niemi:2011ix}. The elliptic flow of charged particles
 can be 
well reproduced (Fig. \ref{fig:alicev2}).
The spectra of identified particles compared to calculations are shown in 
Fig. \ref{fig:alicept}. The agreement is satisfactory. However, 
for  light particles the mean transverse momenta seem too large. 
The HBT radii can be accounted for by the viscous hydrodynamic calculation
(Fig. \ref{fig:hbtlhc}) with  the transverse flow 
generated in the expansion of the fluid.
It shows, that the size and the life-time of the system is well understood in
 dynamical models.

\begin{figure}
  \includegraphics[height=.23\textheight]{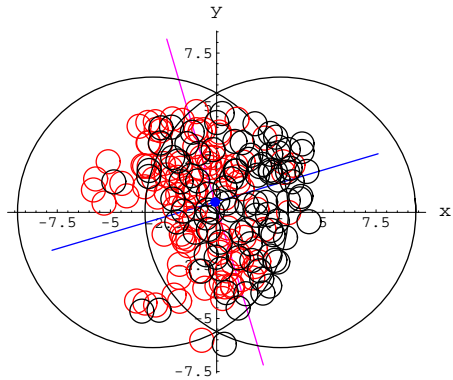}
  \caption{Distribution of wounded nucleons in a Glauber Monte-Carlo event, 
the orientation of the participant plane is rotated with respect to the reaction plane. (from \cite{Broniowski:2007ft})}
\label{fig:glauber}
\end{figure}
The shape of the initial distribution in the transverse plane 
(Fig. \ref{fig:glauber}) is largely influenced by fluctuations
\cite{Andrade:2008xh,Alver:2008zz,Alver:2010gr}.
As a measure of viscosity effects the triangular flow is especially 
important \cite{Alver:2010dn}.
\begin{figure}
  \includegraphics[height=.25\textheight]{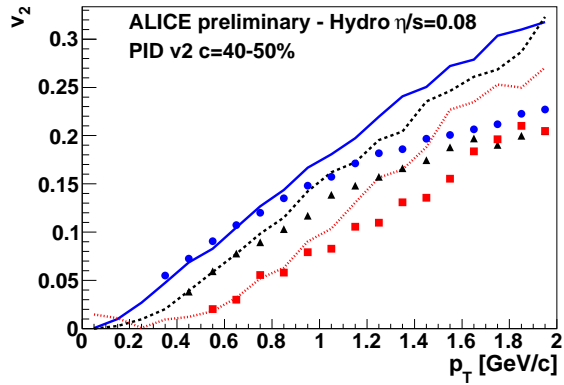}
  \caption{Elliptic flow of $\pi^+$, $K^+$ and $p$
 in Pb-Pb collisions, from hydrodynamic calculations
compared to ALICE Collaboration data \cite{Krzewicki:2011ee}.}
\label{fig:v2}
\end{figure}
In particular, it is difficult for the model to fit at the same time the elliptic and the triangular flow. Another problem, is to reproduce the mass splitting in the $p_\perp$ dependence of $v_2$ between pions and protons.
\begin{figure}
  \includegraphics[height=.25\textheight]{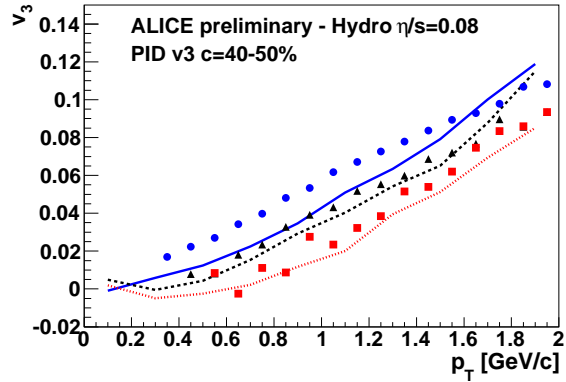}
  \caption{Same as  Fig. \ref{fig:v2} but for triangular flow.}
\label{fig:v3}
\end{figure}
The elliptic flow of identified particles is reasonably well described
 (Fig. \ref{fig:v2}), except for more central events \cite{Krzewicki:2011ee}.
In Fig. \ref{fig:v3}  we illustrate the difficulties of the model 
to reproduce both the strength and the mass splitting of the triangular flow. 
 Some hints of the problems with common predictions for the spectra of pions, 
kaons and protons is already visible in Fig. \ref{fig:alicept}.
It indicates that the very strong transverse collective flow  seen at LHC 
is not entirely accounted for in present hydrodynamic calculations. As noted
 before, the spectra and the 
elliptic flow of identified particles are sensitive 
also to the final hadronic rescattering. More advanced approaches 
to fluctuating initial conditions use event by event hydrodynamic 
calculations \cite{Andrade:2008xh,Schenke:2010rr}.

In summary, the hydrodynamic expansion is a realistic description 
of the behavior of the bulk matter created in relativistic heavy-ion 
collisions. Viscous hydrodynamics reproduces the observed transverse
 momentum spectra, the elliptic flow of charged particles and the HBT radii.
One can conclude that the hot matter created in the collision
behaves like an almost perfect fluid with a cross-over like equation of state.
Fluctuations of the shape of the initial state are evidenced in the 
azimuthal asymmetry of the emitted particles. Some discrepancies remain in the 
description of the momentum dependence of the elliptic an triangular 
flows of identified particles. 

\begin{theacknowledgments}
 The work is supported  by the
Polish Ministry of Science and Higher Education 
grant No.  N N202 263438
\end{theacknowledgments}



\bibliographystyle{aipproc}   

\bibliography{../../hydr}

\IfFileExists{\jobname.bbl}{}
 {\typeout{}
  \typeout{******************************************}
  \typeout{** Please run "bibtex \jobname" to optain}
  \typeout{** the bibliography and then re-run LaTeX}
  \typeout{** twice to fix the references!}
  \typeout{******************************************}
  \typeout{}
 }

\end{document}